\documentclass[12pt]{iopart}
\usepackage{amsmath}
\usepackage{iopams}

\begin{document}

\maketitle

\title{Nonstandard Null Lagrangians and Gauge Functions for Newtonian Law of Inertia}

\author{Z. E. Musielak}
\address{Department of Physics, The University of Texas at 
Arlington, Arlington, TX 76019, USA}
\ead{zmusielak@uta.edu}

\begin{abstract}
New null Lagrangians and gauge functions are derived 
and they are called nonstandard because their forms are different than 
those previously found.  The invariance of the action is used to make 
the Lagrangians and gauge functions exact.  The first exact nonstandard 
null Lagrangian and its gauge function for the law of inertia are obtained,
and their physical implications are discussed.
\end{abstract}


\section{Introduction} 

Most classical and quantum theories of modern physics are formulated
using the standard Lagrangians and Lagrangian formalism [1]. 
There are also nonstandard Lagrangians, whose applications to physics 
are considered in [2], and null Lagrangians, whose relevance to 
ordinary differential equations (ODEs) has already been studied [3].
The main objective of this paper is to derive a new set of nonstandard 
null Lagrangians and their gauge functions, make them exact, and apply 
them to the law of inertia. 

In the calculus of variations, the action $\mathcal{A} [x (t)]$, where 
$x (t)$ is a dynamical variable (or classical object's trajectory) that 
depends on time $t$, is defined as an integral over a local real 
function called a Lagrangian [4], which, for a second-order 
ODE, is denoted as $L (\dot x, x, t)$, where the time derivative 
$\dot x $ represents the particle's velocity in dynamics. The 
Hamilton principle [1,4] 
requires that $\mathcal{A} [x ]$ be stationary (to have either a 
minimum or maximum or saddle point), which is mathematically 
expressed as $\delta \mathcal{A} = 0$, where $\delta$ is the 
functional (Fr\'echet) derivative of $\mathcal{A} [x (t)]$ with 
respect to $x $. The necessary condition that $L (\dot x, x, t)$ 
satisfies the Hamilton principle is $\hat {EL} [ L (\dot x, x, t) ] = 0$, 
where $\hat {EL}$ is the Euler--Lagrange operator [4]. 

A general second-order ODE with constant coefficients is of the
form $\hat D x (t) = 0$, where $\hat D = d^2 / dt^2 + b d / dt 
+ c $ is a linear operator and $b$ and $c$ are constants. Let 
$\hat D_o$ be the operator with  $b = c = 0$, and $\hat D_c$
be the operator with  $b = 0$. Then, for $\hat D_o x (t)  = 0$, 
its Lagrangian depends only on $\dot x^2$.  However, for 
$\hat D_c x (t) = 0$, its Lagrangian depends on both $\dot 
x^2$, the kinetic energy-like term, and $x^2$, the potential 
energy-like term, as originally shown by Lagrange [5]. 
Moreover, if $\hat D x (t) = 0$, its Lagrangian becomes the 
Caldirola--Kanai Lagrangian [6,7]. Thus, Lagrangians 
that depend either on $\dot x^2$ or on $\dot x^2$ and 
$x^2$ are called {\it standard} Lagrangians (SLs). 

Since the standard Lagrangians are not unique, it is also possible 
to construct other Lagrangians that typically depend on $\dot x $ 
and $x $ but not on powers of these variables, and depend also 
on arbitrary functions of the independent variable $t$; Arnold [8] 
refers to such Lagrangians as non-natural Lagrangians.  
Here, Lagrangians that depend on $\dot x $ and $x $, and on 
functions of $t$ are called {\it nonstandard} Lagrangians (NSLs).

The existence of the standard and nonstandard Lagrangians is 
guaranteed by the Helmholtz conditions [9]. The procedure 
of finding these Lagrangians  for given ODEs is called the inverse 
(or Helmholtz) problem of the calculus of  variations [10,11]. 
There are different methods to find the standard [12,13,14,15,16]
and nonstandard [15,16,17,18,19,20] Lagrangians.  Generalized 
nonstandard Lagrangians can also be obtained [21] and applied 
to the ODEs, whose solutions are special functions of mathematical 
physics [22]. Other generalizations of the nonstandard Lagrangians 
 have been applied to the Riccati equation [23] and to a Liénard-type 
nonlinear oscillator [24]. 

In addition to the SLs and NSLs, there is also a family of null (or trivial) 
Lagrangians (NLs) for which the Euler-Lagrange (E-L) equation vanishes 
identically [3].  Another property of these NLs is that they can be 
expressed as the total derivative of a scalar function [3], which is 
called a gauge or gauge function [25,26]; in other words, all NLs
have their corresponding gauge functions (GFs).  The NLs were constructed 
and investigated in mathematics, specifically in Cartan and Lapage forms, 
symmetries of Lagrangians, in Carath\'eodory's theory of fields, and 
extremals and integral invariants [3,27,28,29,30,31,32]. The NLs 
were also applied to elasticity, where they represent the energy density 
function of materials [33,34]. 

The fact that the NLs and their GFs may also play an important role in 
physics was shown recently by using them to restore Galilean invariance 
of the standard Lagrangian for Newton's laws of dynamics [35,36] 
and to add forces to an undriven harmonic oscillator [37].  Since 
those previously constructed NLs resemble the standard Lagrangians, they
are called here {\it standard} NLs, and their corresponding GFs become 
{\it standard} GFs.  In this paper, the standard and nonstandard NLs and 
their GFs are constructed and applied to the law of inertia in Galilean space 
and time with the Galilean group of the metric [8,25].

The outline of the paper is as follows.  In Section 2, previously 
obtained standard null Lagrangians and their gauge functions are described. 
In Section 3, the derived new nonstandard null Lagrangians 
are presented and compared to the standard null Lagrangians.  Invariance 
of the action is used to define the exact nonstandard gauge functions in Section 
4.  Applications of the obtained results to the Newtonian law of 
inertia are presented and discussed in Section 5.  Section 6
summarizes the results of this paper.

\section{Standard Null Lagrangians}    \label{section2}

For the considered ODEs of the form $\hat D x (t) = 0$, its standard Lagrangian 
is given by
\begin{equation}
L_{\rm s} (\dot x , x, t) = {1 \over 2} ( \dot x^2 - c x^2 ) e^{bt}\ . 
\label{S2eq1}
\end{equation}
This Lagrangian was first derived by Caldirola [6] and Kanai [7], 
and it reduces to that given originally by Lagrange [5], if $b=0$.

One of the well-known null Lagrangians [4] is 
\begin{equation}
L_{sn1} (\dot x , x ) = c_1 \dot x  x \ ,
\label{S2eq2}
\end{equation}
where $c_1$ is an arbitrary constant. In this Lagrangian, the power of the 
dependent variables is the same as in the standard Lagrangian given by 
Equation (\ref{S2eq1}); however, the dependent variable and its derivative 
are mixed. 

Recently [35], $L_{sn1} (\dot x , x )$ was generalized to
\begin{equation}
L_{sn2} (\dot x , x , t)  = c_1 \dot x  x  + c_2 ( \dot x  t + x  ) 
+ c_3 \dot x  + c_4\ ,
\label{S2eq3}
\end{equation}
where $c_2$, $c_3$, and $c_4$ are arbitrary constants, and $L_{sn2}$
becomes $L_{sn1}$ if $c_2 = c_3 = c_4 = 0$. This Lagrangian was 
constructed based on the principle that the power of the terms with the 
dependent or independent variable, or their combination, does not 
exceed the power of the terms in the original standard Lagrangian 
given by Equation (\ref{S2eq1}).

Since 
\begin{equation}
L_{sn2} (\dot x , x , t)  = {{d \Phi_{sn2} ( x , t )} \over 
{dt}}\ , 
\label{S2eq4}
\end{equation}
the gauge function $\Phi_{sn2} (x , t) $ is given \cite{35} by
\begin{equation}
\Phi_{sn2} (x , t) = {1 \over 2} c_1 x^2 + c_2 x  t + 
c_3 x  + c_4 t\ .
\label{S2eq5}
\end{equation}

Following [36], the derived $\Phi_{sn2} (x , t)$ is generalized by 
replacing its constant coefficients by arbitrary functions that depend
only on the independent variable. Then, the standard GF, $\Phi_{sn3}  
(x , t)$, can be written as:  
\begin{equation}
\Phi_{sn3} (x , t)  = {1 \over 2} f_1 (t) x^2 + f_2 (t) 
x  t + f_3 (t) x  + f_4 (t) t\ .
\label{S2eq6}
\end{equation}

Since $\Phi_{sn3} (x , t) $ is a function of the variables $x $
and $t$, and its total derivative results in the following general 
standard null Lagrangian
\[
L_{sn3} (\dot x , x , t)  = [ f_1 (t) \dot x  + {1 \over 2}
\dot f_1 (t) x  ] x  + \left [ f_2 (t) \dot x  + \dot f_2 
(t) x  \right ] t 
\]
\begin{equation}
\hskip0.25in + f_2 (t) x   + \left [ f_3 (t) \dot x  + 
\dot f_3 (t) x  \right ] + \left [ f_4 (t) + \dot f_4 (t) t \right ]\ ,
\label{S2eq7}
\end{equation}
where $f_1 (t)$, $f_2 (t)$, $f_3 (t)$, and $f_4 (t)$ are arbitrary but 
at least twice differentiable functions of the independent variable [37].  
Additional constraints on these functions are presented in Section 4, 
where invariance of the action is considered.

The generalization of the gauge function given by Equation (\ref{S2eq6}),
is one natural way to obtain a new NL, but there is also another way,
namely, by replacing the constant coefficients in $L_{sn2} (\dot x , 
x , t)$ (see Equation \eqref{S2eq3}) by the functions $f_1 (t)$, $f_2 (t)$, 
$f_3 (t)$, and $f_4 (t)$. The result is: 
\begin{equation}
L_{sn4} (\dot x , x , t)  = f_1 (t) \dot x  x  + f_2 (t) 
( \dot x  t + x ) + f_3 (t) \dot x  + f_4 (t)\ .
\label{S2eq8}
\end{equation}

Applying $\hat {EL} \{ L_{s,test} (\dot x , x , t)  \} = 0$, 
it is found that $L_{sn4} (\dot x , x , t) $ is a NL if, and only if, 
the following condition 
\begin{equation}
\dot f_1 (t) x  + \dot f_2 (t) t + \dot f_3 (t) = 0\ . 
\label{S2eq9}
\end{equation}
on the functions $f_1 (t)$, $f_2 (t)$, and $f_3 (t)$
is imposed.

There are several different solutions to Equation (\ref{S2eq9}); the 
simplest one is $f_1 (t) = c_1$, $f_2 (t) = c_2$, and $f_3 (t) = c_3$, 
which reduces $L_{sn4} (\dot x , x , t) $ to $L_{sn2} (\dot x , x , 
t)$ without any generalization, but with an additional requirement 
that $f_4 (t) = c_4$. More interesting cases are:  (i) $f_1 (t) 
= c_1$, which gives $ \dot f_2 (t) t = - \dot f_3 (t)$; (ii) $f_2 (t) 
= c_2$ and $ \dot f_1 (t) x  = - \dot f_3 (t)$; and (iii) $f_3 (t) 
= c_3$ with $ \dot f_1 (t) x  = - \dot f_2 (t) t$. In all three 
cases, three new standard NLs are obtained.

Since the functions $f_1 (t)$, $f_2 (t)$, and $f_3 (t)$ in $L_{sn4} 
(\dot x , x , t) $ are limited by the auxiliary condition, given by 
Equation (\ref{S2eq9}), and since $L_{sn3} (\dot x , x , t) $ does 
not require any restrictions on these functions, the standard NL,
given by Equation (\ref{S2eq7}), is more general than 
$L_{sn4} (\dot x , x , t) $; thus, the standard general NL
becomes $L_{sgn} (\dot x , x , t)  = L_{sn3} 
(\dot x , x , t) $.    

The following Corollary summarizes (without a formal proof) 
the results obtained in this Section.

{\bf Corollary:}
The Lagrangian $L_{sgn} (\dot x , x , t) $ is the general null Lagrangian 
among all null Lagrangians that can be constructed
based on the principle that the power of the dependent variable 
cannot exceed the power of this variable in the SL, given by Equation 
(\ref{S2eq1}).

\section{Nonstandard Null Lagrangians}   \label{section3}

Any Lagrangian different from $L_{\rm s} (\dot x , x)$ is a nonstandard
Lagrangian.  Among different known nonstandard Lagrangians, 
the most commonly used [15,16,17,18,19] is: 
\begin{equation}
L_{ns} (\dot x , x , t)  = {{1} \over {g_1 (t) \dot x  +
g_2 (t) x  + g_3 (t)}}\ ,
\label{S3eq1}
\end{equation}
where $g_1 (t)$, $g_2 (t)$, and $g_3 (t)$ are arbitrary and 
differentiable functions to be determined.

Since there are no nonstandard NLs in the literature, the 
objective of this paper is to find them. The procedure is 
based on the two following conditions.  First, for a null 
Lagrangian to be called {\it nonstandard}, it must be of 
different form than the standard NLs, given by 
\mbox{Equations (\ref{S2eq3}) and (\ref{S2eq7}}), and 
its form must be similar to that of Equation (\ref{S3eq1}). 
The latter means that it must contain $\dot x $, $x $, and 
arbitrary functions of $t$, or constants. The second condition 
is similar to that used to 
construct $L_{sn2} (\dot x , x , t) $, $L_{sn3} [\dot x , x , 
t]$ and $L_{sn4} (\dot x , x , t) $, namely, the power of the 
dependent variable and its derivative must not exceed their 
order in the nonstandard Lagrangian given by Equation 
(\ref{S3eq1}). The obtained nonstandard null Lagrangians 
are presented in Propositions 1 and 2, and in the Corollaries 
that follow them.

{\bf Proposition:}
Let $a_1$, $a_2$, $a_3$, and $a_4$ be constants in the following 
nonstandard test-Lagrangian

\begin{equation}
L_{ns,\rm test1} (\dot x , x , t)  = {{a_1 \dot x } \over {a_2 x  + a_3 t 
+ a_4}}\ .
\label{S3eq2}
\end{equation}

Then, $L_{ns,\rm test1} (\dot x , x , t) $ is a null Lagrangian 
if, and only if, $a_3 = 0$.

{\bf Proof:}
Since this Lagrangian must
satisfy the E-L equation, $\hat 
{EL} \{ L_{ns,\rm test1} (\dot x , x , t)  \} = 0$, the required condition 
is $a_1 a_3 = 0$. With $a_1 \neq 0$, then $a_3 = 0$, and $L_{ns,\rm test1}
(\dot x , x , t)  = L_{nsn1} (\dot x , x , t) $, where the latter is
the nonstandard NL. This concludes the proof.

{\bf Corollary:}
Let $L_{nsn1} [\dot x , x ]$ be the nonstandard null Lagrangian 
given 
by: 
\begin{equation}
L_{nsn1} (\dot x , x , t)  = {{a_1 \dot x } \over {a_2 x  + a_4}}\ ,
\label{S3eq3}
\end{equation}
then its gauge function 
$\Phi_{nsn1} ( x )$ is:
\begin{equation}
\Phi_{nsn1} ( x ) = {{a_1} \over {a_2}} \ln \vert a_2 x  + a_4 \vert \ .
\label{S3eq4}
\end{equation}

{\bf Corollary:}
Another nonstandard null Lagrangian that can be constructed is $L_{nsn2} 
(t) = b_1 / ( b_2 t + b_3 )$ and the corresponding gauge function is 
$\Phi_{nsn2} (t) = (b_1 / b_2) \ln \vert b_2 t + b_3 \vert$; however, 
the Lagrangian and gauge function do not obey the first condition; thus, 
they will not be further considered.

Generalization of $L_{nsn1} (\dot x , x)$ is now presented in 
Proposition 2.

{\bf Proposition:}
Let $h_1 (t)$, $h_2 (t)$, and $h_4 (t)$ be at least twice differentiable 
functions and $L_{nsn1} (\dot x , x)$ be the nonstandard null Lagrangian 
given by Equation (\ref{S3eq3}), with the corresponding nonstandard 
gauge function given by Equation (\ref{S3eq4}). A more general 
nonstandard null Lagrangian is obtained if, and only if, the constants 
in $\Phi_{nsn1} ( x )$ are replaced by the functions $h_1 (t)$, $h_2 (t)$, 
and $h_4 (t)$.

{\bf Proof:}
Replacing the constant coefficients $a_1$, $a_2$, and $a_4$ in 
$L_{nsn1} (\dot x , x)$ by the functions $h_1 (t)$, $h_2 (t)$, 
and $h_4 (t)$, respectively, the resulting Lagrangian is:   
\begin{equation}
L_{ns,\rm test2} (\dot x , x , t)  = {{h_1 (t) \dot x } \over {h_2 (t) x  
+ h_4 (t)}}\ .
\label{S3eq5}
\end{equation}

Using $\hat {EL} \{ L_{ns,\rm test2} (\dot x , x , t)  \} = 0$, it is found that 
$L_{ns,\rm test2} (\dot x , x , t) $ is the nonstandard NL only when $h_1 (t) 
= a_1$, $h_2 (t) = a_2$ and $h_4 (t) = a_4$, which reduces $L_{ns,\rm test2} 
(\dot x , x , t) $ to $L_{nsn1} (\dot x , x , t) $ and shows that no
generalization of $L_{nsn1} (\dot x , x , t) $ can be accomplished this 
way.

Now, replacing the constant coefficients in $\Phi_{nsn1} ( x )$ by the 
functions $h_1 (t)$, $h_2 (t)$ and $h_4 (t)$ generalizes the gauge function 
 to 
\begin{equation}
\Phi_{nsgn} ( x ) = {{h_1 (t)} \over {h_2 (t)}} \ln \vert h_2 (t) 
x  + h_4 (t) \vert \ .
\label{S3eq6}
\end{equation}

Since the total derivative of any differentiable scalar function that depends 
on $x $ and $t$ is a null Lagrangian, the following nonstandard NL is 
obtained:

\[
L_{nsgn} (\dot x , x , t)  = {{h_1 (t) [ h_2 (t) \dot x  +
\dot h_2 (t) x ] + \dot h_4 (t)} \over {h_2 (t) [h_2 (t) x  + 
h_4 (t)]}} 
\]
\begin{equation}
\hskip0.5in + \left [ {{\dot h_1 (t)} \over {h_2 (t)}} - {{h_1 
(t) \dot h_2 (t)} \over {h^2_2 (t)}} \right ]\ \ln \vert h_2 (t) x 
+ h_4 (t) \vert\ .
\label{S3eq7}
\end{equation}

As expected, $\hat {EL} \{ L_{nsgn} (\dot x , x , t)  \} = 0$;
thus, $L_{nsgn} (\dot x , x , t) $ is the general nonstandard 
null Lagrangian when compared to Equation \eqref{S3eq3}. 
This concludes the proof.

The derived $L_{nsgn} (\dot x , x , t) $ and $\Phi_{nsgn} (\dot 
x , x , t)$ represent new families of nonstandard general NLs 
and their GFs. These NLs and GFs were derived based on the 
condition that the power of the dependent variable in the 
nonstandard general NLs is either the same as, or lower than, 
that displayed in the original NSL given by Equation (\ref{S3eq1}).

\section{Action Invariance and Conditions for Exactness}  \label{section4}

Since the functions in the standard and nonstandard general null 
Lagrangians are arbitrary, they require either mathematical or physical 
constraints, or both. Among possible mathematical constraints
is invariance of the action, which is used to introduce exact gauge 
functions [36], and symmetries of Lagrangians and the resulting 
dynamical equations [38,39,40,41,42]. Moreover, by using 
Galilean invariance [35], the additional physical constraints 
are imposed on the GFs [36]. Here, only the invariance of 
the action is applied to the obtained standard and nonstandard 
general NLs and GFs; symmetries of Lagrangians are also briefly 
discussed.

In the calculus of variations, the action is defined as:   
\[
A [x ; t_e, t_o] = \int^{t_e}_{t_o} ( L + L_{\rm null} ) dt = 
\int^{t_e}_{t_o} L dt + \int^{t_e}_{t_o} \left [ {{d \Phi_{\rm null} 
(t)} \over {dt}} \right ] dt 
\]
\begin{equation}
\hskip0.25in = \int^{t_e}_{t_o} L dt + [ \Phi_{\rm null} (t_e) - 
\Phi_{\rm null} (t_o)]\ , 
\label{S4eq1}
\end{equation}
where $t_o$ and $t_e$ denote initial and final times, $L$ is a 
Lagrangian that can be either any SL or any NSL, $L_{\rm null}$
is any null Lagrangian and $\Phi_{\rm null}$ is its gauge function.
Since both $\Phi_{\rm null} (t_e)$ and $\Phi_{\rm null} (t_o)$ are 
constants, they do not affect the Hamilton principle that requires 
$\delta A [x ] = 0$. However, the requirement that $\Delta 
\Phi_{\rm null} = \Phi_{\rm null} (t_e) - \Phi_{\rm null} (t_o)$
= const adds this constant to the value of the action. In other 
words, the value of the action is affected by the gauge function.

Using invariance of the action, the following definitions are introduced.

{\bf Definition:}
A null Lagrangian, whose $\Delta \Phi_{\rm null} = 0$, is called 
the {\it exact} null Lagrangian (ENL).

{\bf Definition:}
A gauge function with 
$\Delta \Phi_{\rm null} = 0$
is called the {\it exact} gauge function (EGF).

The condition $\Delta \Phi_{\rm null} = 0$ is satisfied when either 
$\Phi_{\rm null} (t_e) - \Phi_{\rm null} (t_o) = 0$, or $\Phi_{\rm null} 
(t_e) = 0$ and $\Phi_{\rm null} (t_o) = 0$; let the latter be valid. Then, 
the exact null Lagrangians are those whose exact gauge functions make 
the action invariant.

Invariance of the action may now be used to establish constraints on 
the arbitrary functions in the standard NL, $L_{sgn} (\dot x , x , t) $ 
(see Equation \eqref{S2eq7}), and its gauge function, $\Phi_{sgn} (x , t) $ 
(see \eqref{S2eq6}), and make them exact. Taking $\Phi_{sgn} (t_e) 
= 0$ and $\Phi_{sgn} (t_o) = 0$, the following conditions are obtained:
\begin{equation}
{1 \over 2} f_1 (t_e) x_e^2 + f_2 (t_e) x_e t_e + f_3 (t_e) x_e + 
f_4 (t_e) t_e = 0\ ,
\label{S4eq2}
\end{equation}
and 
\begin{equation}
{1 \over 2} f_1 (t_o) x_o^2 + f_2 (t_o) x_o t_o + f_3 (t_o) x_o + 
f_4 (t_o) t_o = 0\ ,
\label{S4eq3}
\end{equation}
with $x_e = x (t_e)$ and  $x_o = x (t_o)$ denoting the end points. If 
the arbitrary functions satisfy these conditions, then $L_{sgn} (\dot x , 
x , t)$ and $\Phi_{sgn} (x , t) $ are exact. The first condition may
be solved by taking $f_3 (t_e) = - f_1 (t_e) x_e / 2$, and $f_4 (t_e) = 
- f_2 (t_e) x_e$. Similar solutions are valid for $t_o$ showing 
that the end values for the functions can be related to each other.

Applying the same procedure to $L_{nsgn} (\dot x , x , t)$, (see 
Equation (\ref{S3eq7})), and to the resulting general gauge function 
$\Phi_{nsgn} (\dot x , x , t)$, (see Equation (\ref{S3eq6})),
the conditions on the arbitrary functions are: 
\begin{equation}
\left [ {{h_1 (t_e)} \over {h_2 (t_e)}} \right ] \ln \vert h_2 (t_e) x_e + 
h_4 (t_e) \vert = 0\ ,
\label{S4eq4}
\end{equation}
and 
\begin{equation}
\left [ {{h_1 (t_o)} \over {h_2 (t_o)}} \right ] \ln \vert h_2 (t_o) x_o + 
h_4 (t_o) \vert = 0\ .
\label{S4eq5}
\end{equation}

Since $\ln [ h_2 (t_e) x  + h_4 (t_e)] \neq 0$ and $\ln [ h_2 (t_e) x  
+ h_4 (t_e)] \neq 0$, both conditions set up stringent limits on the function 
$h_1 (t)$, whose end values must be: $h_1 (t_e) = 0$ and $h_1 (t_o) = 0$; 
however, the procedure does not impose any constraint either on $h_2 (t)$ 
or on $h_4 (t)$.  

Further constraints on all arbitrary functions that appear in the standard 
and non-standard, general, exact null Lagrangians (ENLs) can be imposed 
by considering symmetries of these Lagrangians and the resulting dynamical 
equations. In general, Lagrangians posses less symmetry than the equations 
they generate [38].  Among different symmetries, Noether and non-Noether 
symmetries are identified [39,40,41,42]. The presence of NLs does not 
affect the Noether symmetries [38,41]; however, it may effect the
non-Noether symmetries [42].  All these symmetries impose new 
constraints on the functions.

\section{Applications to Newtonian Law of Inertia}  \label{section5}

Let $(x,y,z)$ be a Cartesian coordinate system, and let $t$ be time 
in all inertial frames; then the one-dimensional motion of a body in one 
inertial frame is given by $\hat D_o x  (t) = 0$, which represents the law 
of inertia. Let $t_o = 0$ and $t_e = 1$ be the end conditions, and let $x (0) 
= x_o = 1$, $x (1) = x_e = 2$ and $\dot x (0) = u_o$ be the initial conditions.  
Then, the solution to $\hat D_o x (t) = 0$ is $x  = u_o t + 1$.  

The {\it standard} Lagrangian for this equation of motion is given by Equation 
(\ref{S2eq1}), with the coefficients $b = c = 0$, and no arbitrary function to 
be determined. However, the standard general NL and the corresponding GF 
are given by Equations (\ref{S2eq7}) and (\ref{S2eq6}), respectively. To 
make the NL and GF exact, the following conditions (see Equations 
(\ref{S4eq2}) and (\ref{S4eq3})) must be imposed on the arbitrary 
functions:
\begin{equation}
f_1 (1) + f_2 (1) + f_3 (1) + f_4 (1) = 0\ ,
\label{S5eq1}
\end{equation}
and 
\begin{equation}
f_3 (0) = - {1 \over 2} f_1 (0)\ .
\label{S5eq2}
\end{equation}

These conditions guarantee that $L_{sgn} (\dot x , x , t) $ 
and $\Phi_{sgn} (x , t) $ are the standard general ENL and 
the standard general  
EGF, respectively.

The {\it nonstandard} Lagrangian
 for the law of inertia is presented by 
the   \linebreak  \mbox{following Proposition.}

{\bf Proposition:}
Let $g_1 (t)$, $g_2 (t)$, and $g_3 (t)$ be arbitrary but differentiable 
functions, and let $\hat D_o x  (t) = 0$ be the equation of motion for 
the law of inertia. Then, the nonstandard Lagrangian for this equation 
of motion is:  
\begin{equation}
L_{ns} (\dot x , x , t)  = {{1} \over {C_1 ( a_o t + v_o)^2}}\
{{1} \over {( a_o t + v_o) \dot x  - a_o x  + C_2}}\ ,
\label{S5eq3}
\end{equation}
where $a_o$, $v_o$, $C_1$, and $C_2$ are constants.

{\bf Proof:}
Following [18], the functions must satisfy
\begin{equation}
{{g_2 (t)} \over {g_1 (t)}} + {1 \over 3} {{\dot g_1 (t)} \over {g_1 (t)}} 
= 0\ ,
\label{S5eq4}
\end{equation}
\begin{equation}
{{\dot g_2 (t)} \over {g_1 (t)}} - {1 \over 2} {{\dot g_1 (t)} \over {g_1 (t)}}
{{g_2 (t)} \over {g_1 (t)}} + {{g^2_2 (t)} \over {2 g^2_1 (t)}} = 0\ ,
\label{S5eq5}
\end{equation}
and 
\begin{equation}
{{\dot g_3 (t)} \over {g_1 (t)}} - {1 \over 2} {{\dot g_1 (t)} \over {g_1 (t)}}
{{g_3 (t)} \over {g_1 (t)}} + {{g_3 (t)} \over {g_1 (t)}} {{g_2 (t)} \over 
{2 g_1 (t)}}= 0\ .
\label{S5eq6}
\end{equation}

Eliminating $g_2 (t)$ from Equations (\ref{S5eq4}) and (\ref{S5eq5}), and 
defining $u (t) = \dot g_1 (t) / g_1 (t)$, one obtains: 
\begin{equation}
\dot u (t) + {1 \over 3} u^2 (t) = 0\ ,
\label{S5eq7}
\end{equation}
which is a special form of the Riccati equation. Following [22], the 
solution to Equation (\ref{S5eq7}) is: 
\begin{equation}
u (t) = 3 {{\dot v (t)} \over {v (t)}}\ ,
\label{S5eq8}
\end{equation}
with $v(t)$ representing a solution to $\ddot v (t) = 0$, which is the 
auxiliary condition [21,22].  

The initial conditions $v (t=0) = v_o$ and $\dot v (t=0) = a_o$ are 
different from those used for $\hat D_o x (t) = 0$. Then, the solution 
becomes $v (t) = a_o t + v_o$, and the functions $g_1 (t)$, $g_2 (t)$
and $g_3 (t)$ become
\begin{equation}
g_1 (t) = C_1 ( a_o t + v_o)^3\ ,
\label{S5eq9}
\end{equation}
where $C_1$ is an integration constant. Having obtained $g_1 (t)$, 
$g_2 (t)$ becomes
\begin{equation}
g_2 (t) = - C_1 a_o ( a_o t + v_o)^2\ .
\label{S5eq10}
\end{equation}
Finally, $g_3 (t)$ can be found by eliminating $g_1 (t)$ and $g_2 (t)$
from Equation (\ref{S5eq5}).  The solution is
\begin{equation}
g_3 (t) = C_1 C_2 ( a_o t + v_o)^2\ ,
\label{S5eq11}
\end{equation}
where $C_2$ is an integration constant.  

Substituting $g_1 (t)$, $g_2 (t)$, and $g_3 (t)$ into Equation (\ref{S3eq1}), 
for $\hat D_o x (t) = 0$, the following final form of the NSL is obtained 
\begin{equation}
L_{ns} (\dot x , x , t)  = {{1} \over {C_1 ( a_o t + v_o)^2}}\
{{1} \over {( a_o t + v_o) \dot x  - a_o x  + C_2}}\ ,
\label{S5eq12}
\end{equation}
which is the same as that given by Equation (\ref{S5eq3}). This concludes 
the proof.

The derived NSL depends on two constants, $a_o$ and $v_o$, 
which are given by the initial conditions for $\ddot v (t) = 0$, 
and two arbitrary constants, $C_1$ and $C_2$, which may be 
determined by the initial conditions for $\hat D_o x  = 0$.
It is easy to verify that $L_{ns} (\dot x , x , t) $ gives 
$\hat D_o x (t)  = 0$ when substituted into the E-L equation. 
This is the first example of the nonstandard Lagrangian for the 
Newtonian law of inertia.

The {nonstandard general} null Lagrangian
$L_{nsgn} (\dot x , 
x , t) $, and its gauge function, 
$\Phi_{nsgn} (x , t) $, are given by Equations
(\ref{S3eq7}) and (\ref{S3eq6}), respectively. To make this 
NL and its GF exact, the following conditions must 
be obeyed (see Equations \ref{S4eq4} and \ref{S4eq5})):
\begin{equation}
\left [ {{h_1 (1)} \over {h_2 (1)}} \right ] \ln \vert 2 h_2 (1) + 
h_4 (1) \vert = 0\ ,
\label{S5eq13}
\end{equation}
and 
\begin{equation}
\left [ {{h_1 (0)} \over {h_2 (0)}} \right ] \ln [ h_2 (0) + h_4 (0)] = 0\ .
\label{S5eq14}
\end{equation}

Since $\ln [ h_2 (1) + h_4 (1)] \neq 0$ and $\ln [ h_2 (0) + h_4 (0)] \neq 0$,
the end values of the function $h_1 (t) $ must be: $h_1 (1) = 0$ and 
$h_1 (0) = 0$, but the end values of either $h_2 (t)$ or $h_4 (t)$ are 
not limited by the conditions for exactness. Further constraints on the 
functions $h_1 (t)$, $h_2 (t)$, and $h_4 (t)$ may be imposed by 
considering symmetries and Lie groups [26] of the derived nonstandard 
general ENL (see Section 4). With these constraints, the first 
nonstandard EGF for the law of inertia is obtained.

The derived nonstandard null Lagrangians are of different forms when 
compared to the standard null Lagrangians obtained in Section 3;
therefore, it is suggested that the NLs be divided into two separate sets. In 
previous work [35,36], it was shown that standard null Lagrangians 
and their gauge functions can be used to restore Galilean invariance of Lagrangians 
in classical mechanics, and to introduce classical forces. The main physical 
implication of the results obtained in this paper is that similar restoration 
of invariance of Lagrangians and definition of forces can also be carried out by 
using the derived general nonstandard NLs, and that the resulting forces will be
of different forms from those previously determined  [36]. The fact 
that not all general standard NLs contribute to the forces was shown by [37]; 
only NLs of special forms can be used to define forces [36,37]. In general, 
most NLs have no influence on these forces. It remains to be determined 
whether the derived general nonstandard null Lagrangian and its gauge function 
define forces, and whether they can be used to convert the first law of dynamics 
into the second law; however, such studies are beyond the scope of this paper.    

The presented methods of finding general standard and nonstandard 
ENLs and their EGFs can be extended to all second-order ODEs of the 
form $\hat D x (t) = 0$, which includes the equations of motion of 
undamped and damped oscillators, and other dynamical systems.  In
previous work [43], the general standard ENLs and EGFs were 
derived for the Bateman oscillators; however, the nonstandard ENLs 
and EGFs are yet to be obtained. The presented methods may also 
be generalized to partial differential equations of quantum mechanics, 
such as the Schr\"odinger equation.

\section{Conclusions}  \label{section6}

This paper presents methods to construct null Lagrangians. Using 
these methods, two different sets of null Lagrangians were obtained 
and classified as standard and nonstandard. The corresponding sets 
of gauge functions were also derived. The presented general standard 
null Lagrangians are known, but the general nonstandard null Lagrangians 
obtained in this paper are new. Since there are differences in the forms 
and properties of the two sets of null Lagrangians, it is suggested that 
these Lagrangians be divided into two classes that correspond to 
these sets.  

The invariance of the action is used to introduce the exactness of both 
general standard and nonstandard gauge functions. Having obtained the 
exact gauge functions, they are used to derive the exact null Lagrangians.
All null Lagrangians and gauge functions, derived in this paper are exact, 
which gives constraints on the end values of the arbitrary functions these
Lagrangians and gauge functions depend on. Further constraints can be 
imposed by symmetries of the exact null Lagrangians and the corresponding 
exact gauge functions, as well as their underlying Lie groups [26].

The obtained results are applied to the ordinary differential equation (ODE)
that represents the law of inertia for which the general exact nonstandard 
null Lagrangian and the corresponding general exact gauge function are 
derived. It is suggested that the derived Lagrangian and gauge function be 
used to restore Galilean invariance of the standard Lagrangian for this law, 
and to introduce forces as carried out in the previous work for the general 
standard null Lagrangians [35,36,37]. Since there are significant 
differences between the general standard and nonstandard null Lagrangians, 
the resulting forces must also be different, which may allow establishing 
a general procedure of defining forces in classical mechanics independently 
from Newton's law of dynamics.

Finally, it must be pointed out that the same method can be used to obtain
the general exact nonstandard null Lagrangians and their gauge functions 
for any ODE given by $\hat D x (t) = 0$, and that it can be extended to 
homogeneous and inhomogeneous partial differential equations, and 
applied to physical problems described by these equations.

\end{document}